\documentclass{llncs}
\usepackage[latin1]{inputenc}
\usepackage[cmex10]{amsmath}
\usepackage{amsfonts}
\usepackage{amssymb}
\usepackage{graphicx}
\usepackage{cite}

\usepackage{color}
\usepackage{enumerate}
\usepackage{pdflscape}
\usepackage{float}

\usepackage{bsymb}
\usepackage{b2latex}

\definecolor{orange}{rgb}{0.9,1.0,0.9}
\definecolor{dgreen}{rgb}{0.0,0.4,0.0}
\definecolor{dblue}{rgb}{0.0,0.0,0.6}
\definecolor{lgray}{rgb}{0.4,0.4,0.4}
\definecolor{gray}{rgb}{0.3,0.3,0.3}

\newcommand{\Snippet}[1] {
	\begin{center}
	\centerline{{\footnotesize{
	$
	\begin{array}{@{}l} #1 \end{array}
	$
	}}}
	\end{center}
}

\begin{document}

\title{Automating Verification of Event-B Models}

\author{Paulius Stankaitis, Alexei Iliasov, \\ David Adjepon-Yamoah, Alexander Romanovsky}

\institute{Centre for Software Reliability, \\ Newcastle University, \\ Newcastle upon Tyne, UK \vspace{0.2cm}\\ \email{ \{paulius.stankaitis, alexei.iliasov,  d.e.adjepon-yamoah, alexander.romanovsky\}@ncl.ac.uk}}

\date{}

\maketitle

\begin{abstract}
Event-B is one of more popular notations for model-based, proof driven specification. It offers a fairly high-level mathematical language based on FOL and ZF set theory and an economical yet expressive modelling notation. Model correctness is established by discharging proving a number conjectures constructed via a syntactic instantiation of schematic conditions. A large proportion of provable conjectures requires proof hints from a user. For larger models this becomes extremely onerous as identical or similar proofs have to be repeated over and over,
especially after model refactoring stages. In the paper we briefly present a new Rodin Platform proof back-end based on the Why3 umbrella prover.
\end{abstract}

\section{Introduction}

Event-B \cite{EventBBook} is one of more popular notations for a model-based, proof driven specification. It offers a fairly high-level mathematical language based on FOL and ZF set theory and an economical yet expressive modelling notation centred around the notion of an atomic event - a form of before-after predicate. Leaving aside the methodological qualities of Event-B, one can regard an Event-B model as a high-level notation from which a number of \emph{proof obligations} may be automatically derived. Proof obligations seek to establish properties like the preservation of invariant and satisfaction of refinement obligations. A model is deemed correct when all proof obligations are successfully discharged. 

Recently some important work has been done to bring a large number of TPTP and SMT-LIB provers under the roof of a common, versatile notation - the Why3 verification platform. At the basic level Why3 offers a common interface to over a dozen of automated provers; it also has its own high-level specification notation to reason about software correctness though we do not make any use of it in this work and rather rely on Why3 to offer a bridge to tools like Z3, SPASS, Vampire and Alt-Ergo.

A theorem prover is a computationally and memory intensive program typically run for rather short periods of time (the vast majority of proofs is done within two seconds) with long idling periods in between. Proof success and perceived usability depend on the capability of an execution platform. Such requirement is best met by the cloud technology.

Doing proofs on a cloud opens possibilities that we believe were previously not explored, outside, perhaps, prover contests. The cloud service keeps a detailed record of each proof attempt along with (possibly obfuscated) proof obligations, supporting lemmas and translation rules. There is a fairly extensive library of Event-B models constructed over the past 15 years and these are a ready of source

\section{Rodin Why3 plug-in}
Development in Event-B is supported by the Rodin Platform \cite{eventBSite} that has been under active development since 2005. It has been long recognised that the Rodin Platform may significantly benefit from an interface between Event-B and TPTP \cite{TPTP} provers. To simplify translation we decided to use the Why3 \cite{boogie11why3} umbrella prover that offers a single and quite palatable input notation and also supports SMT-LIB compliant provers. Why3 supports 16 external automatic provers (not counting different versions of the same tool), these include all the state-of-the-art tools like Z3\cite{Z3}, SPASS\cite{SPASS}, Vampire\cite{Vampire} and Alt-Ergo \cite{AltErgo}.
 
A plug-in to the Rodin Platform was realised \cite{ABZ2016} to map between the Event-B mathematical language and the Why3 \emph{theory} input notation (we do not make use of its other part - a modelling language notation). The syntactic part of the translation is trivial: just one Tom/Java class mapping between Event-B and Why3 operators. The bulk of the effort is in the axioms and lemmas defining the properties of the numerous Event-B set-theoretic constructs. We have a working prototype able to discharge (via provers like SPASS and Alt-Ergo) a number of properties that previously required interactive proof. At the same time, we realise that axiomatisation of a complex mathematical language like the one of Event-B is likely to be an ever open problem. It is apparent that different provers prefer differing styles of operator definitions: some perform better with an inductive style (i.e., to define set cardinality one may say that the size of an empty set is zero, adding one element to a set increases its size by one) while others prefer regress to already known concepts (there exists a bijection such that ...). Since we do not know how to define one best axiomatization, even for any one given prover, we offer an open translator with which a user may define, with as many cross-checks as practically reasonable, a custom embedding of Event-B into the Why3.

\section{Generic lemmata}
Throughout our research we discovered that with the new verification tool we could address another interactive proof problem - fragility and non reusability. There is a number of circumstances when existing interactive proofs become invalidated and a new version of an undischarged proof obligation appears. 

On rare occasions a model or its sizeable part are changed significantly so that there is no or little connection between old and new proof obligations. Far more common are incremental changes that alter the goal, set of hypotheses, identifier names or types. During the refactoring of a refinement tree it is very common to lose a large proportion of manual proofs.

While there is a potential to improve the way the Rodin Platform handles interactive proofs, the fragility of such proofs has mainly to do with their nature. Unlike more traditional theorems and lemmas found in maths textbooks, model proof obligations have no meaning outside of the very narrow model context. And since Event-B relies on syntactic proof rules for invariant and refinement checks, even fairly superficial syntactic changes would result in new proof obligations which are, in fact, if not logically equivalent are often quite similar to the deleted ones.

Even in the case of a significant model change, it is, in our experience, likely that proof obligations similar to those requiring an interactive proof re-appear. In addition, there is a large number of essentially identical interactive proofs re-appearing in different projects due to specific weaknesses in the underlying automatic provers.

%We hypothesise that over time a modeller develops a modelling style characterised by some fairly stable patterns in the usage of the mathematical language and specification constructs. And this leads to stable patterns in the generated proof obligations. For instance, some modellers prefer separate definitions of the direct and converse forms of a relation whilst others construct the converse form on the fly. 

The key to our approach is understanding what 'similar' means in the relation to some two proof obligations. One interpretation is that similar conditions can be discharged by the same proof scripts. To make it practical, this has to be relaxed with some form of a proof script template \cite{FreitasW14}. The interpretation we take in this work is that two proof obligations are similar if they both can be discharged by adding same schematic lemma to the set of their hypotheses. This definition is rather intricately linked with the capabilities of underlying automated provers: adding a tautology (a proven lemma) to hypotheses does not change a conjecture but it might help to guide an automated prover to successful proof completion.

It is our experience that the existing the Rodin automatic provers do not benefit from adding a schematic lemma (with instantiated type variables, to make it first order) to hypotheses and they still need to be instantiated manually by manually by an engineer to have any effect. However, in the case of the Why3 plug-in, with which this approach has a close integration, it is different: a fitting schematic lemma in hypotheses makes proof nearly instantaneous.

There are situations when the only viable way to complete a proof is by providing a proof hint. One such case - refinement of event parameters - is adequately addressed at the modelling notation level where a user is requested to provide a witness as a part of a specification. There are proposals to generalise this, for the majority of situations, and define hints at the model level \cite{ProofHints}.

A schematic lemma considered on its own is of a little use. But if a proof obligation can be proven by adding a schematic lemma, then the construction of a schematic lemma in itself a proof process. As a simple illustration, consider the following (trivial) conjecture:

\Snippet{
\mathit{library} \in \mathrm{BOOKS} \tfun \nat \\
b \in \mathrm{BOOKS} \land c \in \nat \\
\dots \\
\vdash \\
\mathit{library} \ovl \{b \mapsto c\} \in \mathrm{BOOKS} \tfun \nat
}

And suppose there were no automated prover capable of discharge it. It is clear that the crux of the statement is in the interaction of functional override, totality and functionality. The above can be rewritten as

\Snippet{
\mathit{f} \in A \tfun B \\
\vdash \\
\forall x, y \cdot
x \in A \land y \in B \limp
\mathit{f} \ovl \{x \mapsto y\} \in A \tfun B
}

Since the Event-B mathematical language does not have type variables such a condition may only be defined either for specific $A$'s and $B$'s, or, in a slightly altered form, using the Theory plug-in \cite{Butler2013}. But to discharge the original proof obligation one still needs to find this lemma and instantiates it. It is a tedious and error-prone process for a human but a fairly trivial task for a certain kind of automated provers. 

The example above is quite generic in the sense it is potentially useful for in many other contexts. At times a schematic lemma need to be fairly concrete. It is also easier to write a lemma that narrowly targets a proof obligation. This distinction between 'general' and 'specific' is, at the moment, completely subjective and relies on the modeller's intuition. To reflect the fact that a more general lemma is more likely to be reused, schematic lemmas are classified into three visibility classes: machine (single model), project (collection of models) and global. A machine-level lemma will be considered for a proof obligation of the machine with which the lemma is associated; similarly, for the project-level attachment. A global schematic lemma becomes a part of the Event-B mathematical language definition for the Why3 plug-in. 

Just as model construction is often an iterative process, we have discovered during our experiments that finding a good schematic lemma may require several attempts. A common scenario is that an existing lemma may be relaxed so that while it is still strong enough to discharge conditions that were dependent on it, it can also discharge some new ones. For instance, we have seen several cases where a fairly narrow and detailed lemma would gradually slim down to a simple (and much more valuable) statement about distributivity of certain operators. It does require at times a considerable effort to come up with an abstract and minimal covering condition but the result is rewarding and reusable across projects.

\section{Hypotheses and lemmata filtering}

The initial experiments have shown that a minimal axiomatisation support is not sufficient to discharge a sizeable proportion of proof obligations. Provable lemmas were added to assist with specific cases but then it become clear that a large number of support conditions slow down or even preclude a proof. On top of that, the auto tactic language of Rodin offers a very crude hypotheses selection mechanism that for larger models tends to include tens if not hundreds of irrelevant statements. It was thus deemed essential to attempt to filter out unnecessary axiomatisation definitions, Why3 support lemmas and proof obligation hypotheses.

The Rodin mechanism for hypotheses filtering is based on matching conditions with common free identifiers. To complement this mechanism we do filtering on the structure of a formula. It is also a natural choice since support lemmas do not have any free identifiers.

Directly comparing some two formulae is expensive: a straightforward algorithm (tree matching) is quadratic unless memory is not an issue. We use a computationally cheap proxy measure known as the Jaccard similarity which, as the first approximation, is defined as:

\normalsize{
\begin{equation}
JS(P, Q) = \card(P \cap Q) /\card(P \cup Q)
\end{equation}
}

The key is in computing the number of overall and common elements and, in fact, defining what an "element" means for a formula. One immediate issue is that $P$ and $Q$ are sets and a formula, at a syntactic level, is a tree. 

One common way to match some two sequences (e.g., bits of text) using the Jaccard similarity is to use \emph{shingles} of elements to attempt to capture some part of the ordering information. A shingle is a tuple preserving order of original elements but seen as an atomic element. Thus sequence $[a, b, c, d]$ could be characterised by two 3-shingles $P = \{[a, b, c], [b, c, d]\}$ (here $[b, c, d]$ is but a structured name) and matching based on these shingles would correctly show that $[a, b, c, d]$ is much closer to $[a, b, c, d, e]$ than to $[d, c, b, a]$. Trees are slightly more challenging. 

On one hand, a tree may be seen (but not defined uniquely) as a set of paths from a root to leaves and we could just do matching on a set of sequences and aggregate the result. This is not completely satisfactory as tree structure is not accounted for. So we add another characterisation of tree as a set of sequences of the form $[p, c_1, \dots, c_2]$ where $p$ is a parent element and $c_1, \dots, c_2$ are children. This immediately gives a set of $n$-shingles that might need to be converted into shorter $m$-shingles to make things practical.

As an example, consider the following expression $a * (b + c / d) + e * (f - d * 2)$. We are not interested in identifiers and literals so we remove them to obtain tree $+ (* (+ /)) (* (- *))$ which has the following 3-shingles based on paths, $[*, +, /], [+, *, +], [+, *, -], [*, -, *]$, and only 1 3-shingle, $[+, *, *]$, based on the structure.
The shingles are quite cheap to compute (linear to formula size) and match (fixed cost if we disregard low weight shingles, see below). Let $\mathrm{sd}(P)$ and $\mathrm{sw}(P)$ be set of depth and structure shingles of formula $P$. Then the similarity between some $P$ and $Q$ is computed as:

\normalsize{
\begin{equation}
  \renewcommand{\arraystretch}{1.3}
  \begin{array}{ll}
    s(P, Q) = \sum_{i \in I_1} \mathrm{w_d}(i) + c \sum_{i \in I_2} \mathrm{w_w}(i) \vspace{0.3cm} \\ 
    I_1 = \mathrm{sd}(P) \cap \mathrm{sd}(Q), I_2 = \mathrm{sw}(P) \cap \mathrm{sw}(Q)
  \end{array}
\end{equation}
}

where $w_*(i)=\mathrm{cnt}(i)^{-1}$ and $\mathrm{cnt}(i)$ is number of times $i$ occurs in all hypotheses and support lemmas. Very common shingles contribute little to the similarity assessment and may be disregarded so that there is some $k$ such that $\card(I_1) < k, \card(I_2) < k$.

\bibliographystyle{plain}
\bibliography{allrefs2}

\end{document}